\documentclass[12pt,preprint]{aastex}

\def\arcsecpoint{$''\!.$}
\def\deg{$^{\rm o}$}

\def\gtsim{\raisebox{-.5ex}{$\;\stackrel{>}{\sim}\;$}}

\shortauthors{Ruiz et al.}
\shorttitle{Slitless Spectra of Seyfert Galaxies}

\begin{document}

\title{Probing the Kinematics of the Narrow-Line Region in Seyfert Galaxies 
with Slitless Spectroscopy: Observational Results\altaffilmark{1}}

\author{J.R. Ruiz\altaffilmark{2},
D.M. Crenshaw\altaffilmark{3},
S.B. Kraemer\altaffilmark{2},
G.A. Bower\altaffilmark{4},
T.R. Gull\altaffilmark{5},
J.B. Hutchings\altaffilmark{6},
M.E. Kaiser\altaffilmark{7},
\& D. Weistrop\altaffilmark{8}}

\altaffiltext{1}{Based on observations made with the NASA/ESA Hubble Space 
Telescope, obtained at the Space Telescope Science Institute, which is 
operated by the Association of Universities for Research in Astronomy, Inc., 
under NASA contract NAS 5-26555. These observations are associated with 
proposal GO-8340.}

\altaffiltext{2}{Catholic University of America and Laboratory for Astronomy
and Solar Physics, NASA's Goddard Space Flight Center, Code 681,
Greenbelt, MD  20771; ruiz@yancey.gsfc.nasa.gov}

\altaffiltext{3}{Department of Physics and Astronomy, Georgia State 
University, Astronomy Offices, One Park Place South SE, Suite 700,
Atlanta, GA 30303; crenshaw@chara.gsu.edu}

\altaffiltext{4}{Computer Sciences Corporation, Space Telescope Science 
Institute, 3700 San Martin Drive, Baltimore, MD 21218}

\altaffiltext{5}{Laboratory for Astronomy and Solar Physics, NASA's Goddard 
Space Flight Center, Code 681, Greenbelt, MD  20771}

\altaffiltext{6}{Dominion Astrophysical Observatory, National Research Council 
of Canada, 5071 West Saanich Rd., Victoria, BC V9E 2E7, Canada}

\altaffiltext{7}{Department of Physics \& Astronomy, Johns Hopkins University,
3400 North Charles St., Baltimore, MD 21218}

\altaffiltext{8}{Department of Physics, University of Nevada at Las Vegas, 
4505 Maryland Parkway, Las Vegas, NV 89154-4002}

\begin{abstract}

We present slitless spectra of 10 Seyfert galaxies observed with the Space 
Telescope Imaging Spectrograph on the {\it Hubble Space Telescope} ({\it HST}). 
The spectra cover the [O~III] $\lambda\lambda$4959, 5007 emission lines at a 
spectral resolving power of $\lambda$/$\Delta\lambda$ $\approx$ 9000 and a 
spatial resolution of $\sim$0\arcsecpoint1. We compare the slitless spectra 
with previous {\it HST} narrow-band images to determine the velocity shifts and 
dispersions of the bright emission-line knots in the narrow-line regions (NLRs) 
of these Seyferts, which extend out to at least several hundred pc from their 
nuclei. Many knots are spatially resolved with sizes of tenths of 
arcsecs, corresponding to tens of pcs, and yet they appear to move coherently 
with radial velocities between zero and $\pm$ 1200 km s$^{-1}$ with respect to 
the systemic velocities of their host galaxies. The knots also show a broad 
range in velocity dispersion, ranging from $\sim$30 km s$^{-1}$ (the velocity 
resolution) to $\sim$1000 km s$^{-1}$ FWHM. 
Most of the Seyfert galaxies in this sample show an organized flow pattern, 
with radial velocities near zero at the nucleus (defined by the optical 
continuum peak) and increasing to maximum blueshifts and redshifts within 
$\sim$ 1$''$ of the nucleus, followed by a decline to the systemic velocity.
However, there are large local variations around this pattern and in one case 
(NGC~7212), the radial velocities are nearly chaotic. The emission-line knots 
also 
follow a general trend of decreasing velocity 
dispersion with increasing distance from the nucleus. In the Seyfert 2 
galaxies, the presence of blueshifts and redshifts 
on either side of the nucleus indicates that rotation alone cannot 
explain the observed radial velocities. The most straightforward 
interpretation is that radial outflow plays an important 
role in the NLR kinematics. Each of the Seyfert 
galaxies in this sample (with the exception of 
Mrk 3) shows a bright, compact (FWHM $\leq$ 0\arcsecpoint5) [O III] knot 
at the position of its optical nucleus. These nuclear emission-line knots have 
radial-velocity centroids near zero, but they typically have the highest 
velocity dispersions. Their similar properties suggest they may be a common, 
distinct component of the NLR.

\end{abstract}

\keywords{galaxies: Seyfert}
~~~~~

\section{Introduction}

The narrow-line region (NLR) in Seyfert galaxies is the smallest resolvable 
structure in the UV and optical that is directly affected by the ionizing 
radiation and dynamical forces from the inner active nucleus. Hence, studies of 
the physical conditions and kinematics of the NLR should yield important 
insights into 
the machinery of active galactic nuclei (AGN), as material is processed into 
and out of the environment surrounding the central supermassive black 
hole (SMBH). The NLR also provides information on the interactions
between the host galaxy and the inner unresolved nucleus, which is thought to 
include a central SMBH, accretion disk, broad-line region (BLR), and dusty 
molecular torus (see Jaffe et al. 2004). By definition, the NLR is responsible 
for narrow emission 
lines with widths $\approx$ 500 km s$^{-1}$ full-width at half-maximum (FWHM), 
whereas emission lines from the BLR have widths in the range 1000 -- 
7000 km s$^{-1}$ FWHM. Also by definition, Seyfert 2 galaxies show only narrow 
emission lines in their UV and optical spectra, while Seyfert 1 galaxies show 
both broad and narrow emission lines.
Whereas the BLR has a size of only tens of light days (Peterson et al. 2004), 
the NLR extends out to at least several hundred pcs from the nucleus (Schmitt 
et al. 2003a, b). Nevertheless, the angular sizes of the NLRs in nearby Seyfert 
galaxies are still quite small; for example, at a redshift of 
0.01 and H$_0$ $=$ 75 km s$^{-1}$ Mpc$^{-1}$ , 100 pc subtends only 
$\sim$0\arcsecpoint5.
Thus, detailed studies of the physical conditions and kinematics across the NLR 
were not possible before the advent of the {\it Hubble Space Telescope} ({\it 
HST}) and ground-based adaptive optics.

{\it HST} has provided emission-line images of the NLRs in many nearby Seyfert 
galaxies at a spatial resolution of $\sim$0\arcsecpoint1. The images show that 
NLRs tend to be clumpy, with many resolved ``knots'' of emission that 
are typically a few tenths of an arcsec in size (Evans et al. 1991, 1993; 
Schmitt et al. 2003a). The overall morphology of the NLR is varied; 
triangular (or ``fan-like''), linear, spiral (or ``S'' shaped), elliptical, 
circular, 
arc-like, and amorphous shapes are all
common. However, there is a trend in the sense that the NLRs in Seyfert 2 
galaxies are more elongated than those in 
Seyfert 1 galaxies, which tend to be more concentrated and circular in 
appearance (Schmitt et al. 2003b)
This result is in 
qualitative agreement with the unified model (Antonucci 1993), which posits 
that the two types of Seyfert galaxies are due to our 
viewing angle with respect to an optically thick torus outside of the BLR, 
which produces bicones of ionizing radiation in the NLR. 
In the case of a Seyfert 2 galaxy, the torus blocks our view of the BLR, and we 
see a more extended, triangular NLR. For a Seyfert 1 galaxy, we are 
looking down the cone to see the BLR, and the NLR looks more compact and 
circular\footnote{Two notable exceptions are the Seyfert 1 galaxies NGC 4151 
(Evans et al. 1993) and NGC 3516 (Ferruit et al. 1998), which show elongated 
NLRs and are likely viewed at intermediate angles.}.
The triangular-shaped NLRs seen in some Seyfert 2 galaxies are clear 
examples of projected bicones (with one of the cones diminished or extinguished 
in some cases by the host galactic plane); other NLR morphologies could 
be explained by selective filling of the ionization bicones by material in ways 
that are admittedly not well understood.

The physical conditions in the NLR have been studied extensively, 
and it is clear that photoionization by the central continuum source 
(the accretion disk) plays a dominant role in heating the gas, although shocks 
may play an important role in localized regions (Kraemer \& Crenshaw 2000; 
Dopita et al. 2002). The 
dynamics of the gas in the NLR, however, are less certain. Prior to {\it HST}, 
most studies had to rely on spatially-integrated line profiles. Unfortunately, 
the profiles can be reproduced by a wide variety of kinematic models (e.g., 
infall, outflow, rotation, parabolic orbits, etc., (e.g., Osterbrock \& 
Mathews, 1986, and references therein; De Robertis \& Shaw 1990; Veilleux 1991; 
Moore \& Cohen 1996). Thus, spatially-resolved spectra across the NLR are 
essential for further kinematic constraints. Early attempts 
with {\it HST} made use of the long-slit spectroscopy mode of the Faint Object 
Camera (FOC), resulting in kinematic information on the NLRs in NGC~4151 (Winge 
et al. 1997; 1999), NGC~1068 (Axon et al. 1998), and Mrk~3 (Capetti et al. 
1999). Several models were advocated based on these data, including 
rotation around the SMBH and interior mass, radial acceleration by radio jets, 
and expansion of ``hot cocoons'' of gas around the jets, or more accurately, 
radio lobes, forcing the NLR clouds away from the lobes (Axon et al. 1998; 
Capetti et al. 1999).

The installation of the Space Telescope Imaging Spectrograph (STIS) on {\it 
HST} in 1997 February provided an ideal opportunity for detailed kinematic 
mapping of the NLR. Using STIS long-slit and slitless observations, we 
presented evidence that radial outflow is the dominant component of motion in 
the NLRs of three Seyfert galaxies: NGC 4151 (Hutchings et al.  1998; Kaiser et 
al. 2000; Nelson et al. 2000; Crenshaw et al. 2000), NGC 1068 (Crenshaw \& 
Kraemer 2000), and Mrk 3 (Ruiz et al. 2001). We found no connection between the 
kinematics of the NLR clouds and the radio lobes (Kaiser et al. 2000; Ruiz et 
al. 2001), and concluded that the outflow was more likely the result of 
radiation pressure or winds than radio jets. Comparison 
of the radial-velocity fields in these three Seyferts with 
biconical outflow models indicated that the NLR gas accelerates outward from 
the nucleus to a turnover point, at a distance of $\sim$100 pc from the 
nucleus, 
and subsequently appears to decelerate to the systemic velocity at a distance 
of $\sim$300 pc.
At larger distances, the velocities are consistent with 
rotation in the galactic plane, which agrees with results from earlier 
ground-based studies of extended narrow-line regions (ENLRs), located at 
distances greater than a few hundred parsecs from the nucleus (Unger et al. 
1987). 

To expand the collection of Seyfert galaxies with NLR kinematic data, we have 
obtained STIS slitless spectra of 10 Seyfert galaxies at high spectral 
resolution. The principal advantage 
of the slitless technique is that it is very efficient: the radial velocities 
of all of the bright knots in the NLR can be mapped with a single pointing of 
{\it HST}. The principal disadvantage, compared to long-slit observations, is 
that faint knots or more diffuse emission cannot be detected and measured with 
this technique. Thus, the sample is biased, in the sense that we picked 
Seyfert galaxies with NLRs that are known to be extended and have high-contrast 
knots, based on existing {\it HST} [O~III] images. In this paper, we present 
the observational results for 
our slitless program. Some of the results for Mrk 3 have already been published 
(Ruiz et al. 2001); we include it in this paper for completeness. In a 
subsequent paper, we discuss our comparison of the observations with 
kinematic models (D.M. Crenshaw et al. 2004, in preparation).

\section{Observations and Analysis}

\subsection{Sample and Technique}

The basic idea behind the slitless technique is to measure the displacement of 
an emission-line knot, originally detected in a narrow-band image, after a 
grating is introduced. The displacement in the dispersion direction yields the 
average radial velocity of the knot, and the broadening of the emission gives 
the velocity dispersion of the knot. To further increase the efficiency of 
this program, we chose Seyfert galaxies with existing {\it HST} [O~III] images, 
which are needed to determine the spatial locations of the 
knots on the sky. We have used this technique successfully in the past, with 
STIS slitless spectra of NGC~4151 (Hutchings et al. 1999; Kaiser et al. 2000). 
To demonstrate the reliability of this technique, we compared
the radial velocities from STIS slitless and long-slit observations 
of NGC~4151 (Crenshaw et al. 2000), and found they differ by only $\pm$50 km 
s$^{-1}$ on average, whereas the range of radial velocities for the bright 
knots is on the order of $\pm$800  km s$^{-1}$.

Table 1 lists our sample, which includes eight Seyfert 2, one Seyfert 1.9 
(Osterbrock 1981), and one Seyfert 1 galaxy. The table also gives the host 
galaxy types, redshifts of the host galaxies (from stellar absorption lines 
when H~I 21-cm observations are not available), and our adjustments to the 
redshifts based on measured ENLR velocities, as described in \S2.3. 
Previous {\it HST} narrow-band images showed that all of these NLRs have 
extended, high-contrast knots of [O III] emission.

We obtained new observations of each Seyfert galaxy with the STIS CCD detector, 
G430M grating, and an open aperture (52$''$ $\times$ 52$''$). The CCD is a 1024 
$\times$ 1024 pixel detector with a plate scale of 0\arcsecpoint051 
pixel$^{-1}$ and an angular resolution of $\sim$0\arcsecpoint1 FWHM in imaging 
mode. With the G430M grating in place, the scale is changed to 0\arcsecpoint057 
pixel$^{-1}$ in the dispersion 
direction due to anamorphic magnification. For a point source, the dispersion 
of the G430M grating is 0.28 \AA\ pixel$^{-1}$, and the spectral resolution is 
twice this value, which yields a resolving power of $\sim$9000 and a 
velocity resolution of $\sim$33 km s$^{-1}$ (FWHM) in the vicinity of the 
[O~III] $\lambda$5007 line. For the extended emission-line knots, the 
resolution is 
degraded, but most have angular sizes $\leq$ 0\arcsecpoint3 and velocity widths 
$>$ 100 km s$^{-1}$ FWHM, so this has little effect on our measurements. Since 
the NLRs are all 
much smaller than the CCD field of view and we did not use a slit, we obtained 
the observations without target acquisition, to maximize the exposure time in 
an orbit. We observed each target for a single orbit.

Table 2 gives a log of our STIS grating observations, which includes the STScI
dataset names, UT dates of observation, central wavelength settings, total 
exposure times, and position angles (from north in an eastward direction) of 
the Y-axis of the CCD chip (which is perpendicular to the dispersion 
direction). The 
wavelength coverage of the G430M grating is only 286 \AA, so we chose the 
central wavelength such that the redshifted [O~III] $\lambda$5007 line 
was well covered. This resulted in coverage of the [O~III] 
$\lambda$4959 line in all cases, and the H$\beta$ line in five cases. The 
listed exposure time is the sum from two separate, equal exposures obtained in 
the orbit for the 
purpose of eliminating cosmic-ray hits. The position angle was chosen to place 
the minor axis of the NLR emission roughly 
along the STIS dispersion axis, to minimize confusion from overlapping 
clouds in the dispersion direction.

To measure the radial velocities of the emission-line knots, we used four 
images of each Seyfert galaxy: a direct continuum and an emission-line image 
from the STScI/MAST archives, a STIS direct broad-band image, and the STIS 
spectral image. In addition, we obtained two calibration images at each 
pointing, as discussed below.
Ideally, one would like to determine the radial velocities of the knots in a 
slitless spectrum by comparing it with a direct [O~III] image obtained with the 
same instrument and orientation. However, given the redshifts of these Seyfert 
galaxies, STIS does not have suitable [O~III] filters for this purpose. Thus, 
we relied on archival {\it HST} emission-line and continuum images obtained 
with the WFPC2, FOC, and WF/PC (PC mode) cameras, in order of preference. 
However, we still required a direct 
STIS image to properly register the archival and STIS spectral images, since 
most of our targets are Seyfert 2 galaxies, and lack a bright central 
point-source (like that in NGC 4151), which is convenient for alignment 
purposes. Thus, we obtained a short (20 s) direct exposure in STIS camera mode, 
which is sensitive from 2000 -- 10,000 \AA\ and is dominated by continuum 
emission, but contains line emission as well. We rotated the archival 
continuum image using the position angle in the header and aligned it with 
the STIS direct image using common features, particularly the 
central continuum peak (due to the AGN in NGC 3516 and stellar emission in the 
Seyfert 1.9 and 2 galaxies). This
also resulted in alignment of the archival [O~III] image with the direct 
STIS image, since it was obtained during the same pointing as the archival 
continuum image. In a number of instances, we were also able to check our 
alignment by comparing emission-line knots detected in the STIS direct image 
with those in the archival [O III] images.

A final translation of all of the direct 
images was needed in the STIS dispersion direction to align them with the STIS 
spectral image,  by placing the emission-line 
knots at their expected locations if they had zero radial velocity with respect 
to the systemic redshift of the host galaxy. This was done with the help 
of two calibration images obtained at each pointing: a direct image of a 
continuum lamp through a 52$''$ $\times$ 0\arcsecpoint1 slit, and a wavelength 
calibration image of an emission-line lamp with the G430M grating and the same 
slit. The displacement in the dispersion direction between the continuum-lamp 
image and the expected position of redshifted [O~III] $\lambda$5007 (from the 
redshifts in Table 1) gives the shift needed to align the direct and spectral 
images, such that the measured displacement for a knot gives the radial 
velocity with respect to systemic. This final step is an improvement over the 
procedure used for NGC~4151 (Kaiser et al. 2000), in that we are able to 
determine absolute radial velocities as opposed to relative values (in that 
case, with respect to the bright central knot of [O~III] emission). The final 
alignment, rotation to the position angle of the STIS images, rebinning to the 
STIS plate scale, and translation in the dispersion direction were actually 
accomplished with a single resampling of the archival images.

Table 3 gives the archival images used for this study. Most of these images 
have been discussed in detail by other authors. The emission-line (narrow-band) 
images are those with an ``N'' suffix, whereas the continuum images (medium- or 
wide-band) images are those with ``M'' or ``W'' suffixes. The FOC narrow-band 
image of Mrk 3 is shown in Ruiz et al. (2001), and was useful for detecting 
extended emission. Mrk 573 has continuum and emission-line images obtained at 
different times, but we were able to align these images using common features.

\subsection{Direct and Dispersed Images}

We use the images of Mrk~3 to illustrate the general effects of the filters 
and grating used for this study.
In Figure 1, we show three archival images of Mrk 3, along with the STIS 
spectral image on the same scale. In the F502N image, the [O III] emission 
dominates in the form of bright knots of emission in a backwards ``S'' shape. 
However stellar continuum emission is also clearly seen in the form of a 
compact circular shape, which dominates the emission in the F547M continuum 
image. The 
F606W filter is broad enough to contain substantial emission from both 
continuum and emission-lines, and deep enough to show some of the extended 
[O~III] emission in a biconical shape. The STIS slitless spectrum shows all of 
these features: the bright central knots show large displacements due to their 
radial velocities and are highly broadened as a result of their velocity 
dispersions. The continuum is faintly visible as a broad horizontal band;
the clouds in the ENLR, at distances $\gtsim$ 1\arcsecpoint5 from the nucleus, 
have substantially lower radial velocities and velocity widths than those in 
the NLR (see also Ruiz et al. 2001).

Figure 2 displays side-by-side [O~III] and STIS slitless images of the 
remaining nine Seyfert galaxies in our sample. The [O~III] images are not 
corrected for 
continuum emission, in order to show the effects of this emission in the STIS 
spectra. Distortions in the spectral images, compared to the direct images, are 
due to radial-velocity shifts and velocity broadening of the emission-line 
knots. The velocity broadening is most apparent in the inner NLRs, typically 
within $\sim$1$''$ of 
the optical continuum peaks. It is clear from the images that most of the
emission-line clouds seen in the direct images can be identified without 
difficulty in the spectral images.

The broad horizontal bands in the STIS spectra near the galaxy nuclei, 
extending over 1$''$ -- 5$''$, are 
primarily due to stellar continuum emission in the host galaxy (the narrow 
bands offset from the nuclei are individual stars in the STIS field of view). 
Most of the structures in the continuum
bands are due to dust lanes, which can be seen in the archival continuum 
images (e.g., Malkan, Gorjian, \& Tam 1998). We have checked for scattered 
light from the hidden nucleus, by 
extracting regions of enhanced continuum emission and looking for broad 
H$\beta$ emission in the five Seyfert 2 galaxies with coverage of this line 
(Mrk 620, NGC 1386, NGC 5506, NGC 5643, and NGC 7212). No broad H$\beta$ 
emission was detected. However, slitless spectra are not optimal for detecting 
scattering regions, since their emission would be convolved with other regions 
of pure stellar emission in the dispersion direction. Thus we cannot rule out 
local regions of scattered emission in these five Seyfert galaxies, such as 
those detected in STIS long-slit spectra of NGC 1068 on scales of 
$\sim$0\arcsecpoint3 (Crenshaw \& Kraemer 2000).

Perhaps the most striking feature in the images in Figure 2 is the bright, 
central knot of [O~III] emission in each object, which is highly dispersed in 
the STIS images. In each case, it is located at the optical continuum peak 
(indicated by a ``+''). These knots are spatially resolved in most cases, with 
sizes that range from 0\arcsecpoint15 to  0\arcsecpoint5 FWHM perpendicular to 
the STIS dispersion. As shown in \S2.3, the
nuclear knots typically have the highest velocity dispersions, but have 
radial-velocity centroids that tend to be close to zero. 

There are a few images that deserve further comment at this point. The FR533N 
image of Mrk 620 shows an extended ring of emission that is not apparent in the 
STIS [O~III] line, but is clearly seen in the STIS H$\beta$ line. The high
H$\beta$/[O~III] ratio in this ring indicates low-ionization gas that is likely
due to starburst activity. However, the small V-shaped feature extending 
toward the east is [O~III] emission from higher-ionization gas that is
presumably directly ionized by the AGN. The F501N image of NGC 5506 only 
covers one side of the NLR (a clear example of a cone in projection), whereas 
the STIS spectral image shows a faint counter-cone south of the optical 
continuum peak; we cannot measure the radial velocities of the knots in this 
region without a comparison image. There is no narrow-band image of NGC 5728,
but the medium-band F492M image shown in Figure 2c contains [O~III] emission
that we can use for registration of the emission-line knots.  This filter
transmits substantially more continuum than the narrow-band filters, which
explains the elliptical disk of continuum emission that shows up as a broad
horizontal band in the STIS spectrum.

\subsection{Measurements}

We processed the STIS spectra with the IDL software developed at NASA's Goddard 
Space Flight Center for the Instrument Definition Team. We removed cosmic-ray 
hits using the two exposures of each target, and removed hot or warm 
pixels by interpolation in the dispersion direction. We used the calibration 
exposures described in \S2.1 to obtain a wavelength scale and processed the 
spectral images in ``extended'' mode, which geometrically rectifies the images 
to produce a constant wavelength along each column and fluxes at each position
in units of ergs s$^{-1}$ cm$^{-2}$ \AA$^{-1}$ per pixel (fluxes are conserved 
in the geometric transformation).

The challenge of slitless spectroscopy is to identify an emission-line
knot, displaced by the introduction of a grating, in both the direct and 
dispersed images. As a guide, we utilized the fact that a given knot's location 
and extent perpendicular to the dispersion are the same in the two images.  
Once we identified the knots, we fit the spatial 
and spectral images row-by-row (parallel to the dispersion and separated by 
0\arcsecpoint05 intervals) with a Gaussian for each knot plus a straight line 
for the background emission. 
We then determined the radial velocity of each knot by subtracting 
the positions of the Gaussian peaks in the direct and dispersed images, and the 
intrinsic velocity width by subtracting the FWHMs of the two Gaussians in 
quadrature. In the few instances where the FWHM was slightly smaller in the 
dispersed image, due to statistical fluctuations, we set the intrinsic 
FWHM to zero. We did not average together values in different rows, so we have 
multiple measurements of each knot, the number of which depends on its extent 
in the spatial direction of the dispersed image. Occasionally, a single 
knot in the direct image would split into two knots in the dispersed image, 
indicating two knots with different radial velocities that appear as one in 
projection.

After determining the radial velocities as a function of position, it became 
evident that some of the galaxies showed systematic offsets from zero in the 
ENLR on both sides of the nucleus. Given the available evidence that the ENLR 
lies in the host galaxy, we added a constant to the radial velocities of 
galaxies for which the average offset was $>$ 50 km s$^{-1}$ (the 
approximate measurement error) to bring the average radial velocity in the ENLR 
to zero. In some cases, there are still asymmetries in the ENLR velocities on 
either side of the nucleus, which can be ascribed to galactic rotation. The 
velocity offsets are given in Table 1. The 
systematic errors in radial velocity could be the 
result of small misregistrations among the various images and/or inaccurate 
values for the systemic velocities. The latter might explain why Mrk 573 has 
such a large 
correction; it is one of only two galaxies for which we had to rely on 
stellar absorption lines for its systemic redshift.

\section{Results}

Most of our analysis focuses on radial velocities and FHWMs as functions of 
projected angular distance from the optical continuum peak (which we refer 
to as the ``nucleus''). The slitless spectra can also be used to look for 
variations in these quantities as a function of distance from the major axis of 
the NLR, or as a function of position angle with respect to the major axis, but 
we found no convincing evidence for such variations. It is possible that they 
are present, but are overwhelmed by strong changes in the velocities as a 
function of radial distance from the nucleus. Since our coverage is limited to 
only the bright knots of emission, more detailed maps of the NLR 
kinematics, using multiple slit positions at high spectral resolution (e.g., 
Cecil et al. 2002), are required to investigate the above, more subtle, effects.

In Figure 3, we show the measured radial velocities and FWHMs for Mrk 3 as a 
function of projected angular distance from the nucleus on the sky. The trend 
in radial 
velocities agrees well with our previous measurements from long-slit, 
low-resolution spectra and from the slitless spectrum in which we averaged the 
velocities for each knot (Ruiz et al. 2001). In fact, individual knots can be 
identified as points clumped together in the radial-velocity plot. 
The advantage of long-slit observations can be seen in Figure 8 of Ruiz et al. 
(2001), which shows a much cleaner trend due to the ability to measure fainter 
emission along the slit. 
The overall trend is an increase in radial velocity from near zero at the 
nucleus to maximum blueshifts and redshifts at $\sim$0\arcsecpoint3 on either 
side of the nucleus, followed by a decline to the systemic velocity (with large 
scatter in Figure 3) at distances $\gtsim$2$''$. The FWHM plot in Figure 3
shows large scatter at each location as well, but there is a trend of higher 
velocity dispersions for the bright inner clouds, as can be seen in Figure 1.

We show radial velocity and FWHM plots for the other nine Seyfert galaxies in 
Figure 4. Row-by-row measurements of each emission-line knot results in radial 
velocities and FWHM that are consistent to within $\pm$50 km s$^{-1}$. 
Furthermore, independent measurements of the same knots in both long-slit and 
slitless spectra yield values that agree to within $\pm$50 km s$^{-1}$ (\S2.1), 
so we adopt this as our typical uncertainity for both radial velocity and FWHM. 
The knots, with angular sizes of tenths of arcsecs, corresponding to tens of 
pcs, move coherently with their own peculiar motions, often superimposed on a 
more general flow. Although the amplitudes of the velocity curves differ 
significantly, there are a number of general trends.

Most of the Seyfert 1.9 and 2 galaxies in the sample show blueshifts and 
redshifts on either side of the nucleus, indicating that simple rotation is not 
the major source of the velocities. The only Seyfert 1 galaxy in the sample, 
NGC 3516, shows primarily blueshifts on one side of the nucleus and redshifts 
on the other side, like the Seyfert 1 galaxy NGC 4151 (Crenshaw et al. 2000). 
Taken together, these results support the biconical outflow picture, in which 
the bicone axis is near the plane of the sky for the Seyfert 2 galaxies and at 
an intermediate angle ($\sim$45\deg) for the two Seyfert 1 galaxies with 
elongated emission, which results in one cone with approaching and the other 
with receding knots. One 
problem with this simple interpretation is that NGC 3516 shows ``rogue'' clouds 
moving in directions opposite to the general flow at 6$''$ -- 7$''$ from the 
nucleus, similar 
to that seen for a few low-intensity, high-velocity clouds in NGC 4151 
(Hutchings et al. 1999; A. Das et al. 2004, in preparation). Another 
interesting case is the Seyfert 2 galaxy 
NGC~5728, which shows blueshifts on one side of its nucleus and redshifts on 
the other side. However, this could be matched with a biconical outflow model 
which is tilted significantly out of the plane of the sky, but not enough to 
allow us to see down the cone to the BLR.

The overall flow pattern in most objects is an increasing radial velocity from 
near zero at the nucleus to maximum blueshifts or redshifts within 
0\arcsecpoint3 -- 2$''$ 
of the nucleus, followed by a more gradual decline to the systemic velocity. 
We have seen this trend before in NGC 1068 (Crenshaw \& Kraemer 2000) and NGC 
4151 (Crenshaw et al. 2000, and reference therein). There are a few exceptions 
to this general trend. 
NGC 1386 has large blueshifts and redshifts at the nucleus and an unusual 
overall flow pattern. NGC 5728 has redshifted velocities at the nucleus, but 
it would fit the general trend if the center position were moved 
$\sim$0\arcsecpoint5 to the west, coincident with the dust lane just above the 
nucleus in Figure 2. In Figure 4, NGC 7212 shows a mostly chaotic distribution 
of velocities with no clear pattern.

The plots in Figure 4 show an overall trend of decreasing FWHM with increasing 
distance from the nucleus, although in most cases there is a wide spread of 
values at each position. The bright nuclear knot of emission seen in every 
Seyfert galaxy in this sample except Mrk 3 has a large, and often the largest, 
FHWM, like that in NGC 4151 (Kaiser et al. 2000). The highest FWHMs tend to be 
comparable to the largest radial velocities in a Seyfert galaxy.
On the other hand, Mrk 3 has several very bright knots of emission with large 
FWHMs near its nucleus, like NGC 
1068 (Cecil et al. 2002). In these two Seyfert 2 galaxies, it may be that we 
are seeing expanded versions of the nuclear knots, since their central 
knots show large velocity dispersions and approximately zero radial velocity 
when averaged together (Crenshaw \& Kraemer 2000; Ruiz et al. 2001).

The nuclear emission-line knots in the majority of Seyfert galaxies 
consist of multiple kinematic components; they are not resolved spatially, but 
can be 
seen in spectral extractions around the nucleus. As an example, we show the 
nuclear spectrum of NGC 3081 in Figure 5. The extraction height perpendicular 
to the dispersion is 0\arcsecpoint15, corresponding to the FWHM of the nuclear 
knot in the spatial direction. There are clearly two major kinematic components 
in the [O~III] $\lambda$5007 line at $\sim$ $-$250 and $\sim$ $+$ 50 km 
s$^{-1}$.

\section{Conclusions}

{\it HST}/STIS slitless spectroscopy of nearby 
Seyfert galaxies provides a useful snapshot of their NLR kinematics. The 
technique works when the NLR is clumpy and extended, so that individual 
emission-line knots can be identified in both direct and dispersed images.
Our sample is therefore biased against NLRs that are smooth and/or compact, 
such as those typically found in Seyfert 1 galaxies (Schmitt et al. 2003b). The 
two Seyfert 1 galaxies that we have successfully mapped with slitless spectra, 
NGC~3516 and NGC~4151, have elongated NLRs. The disadvantage of the slitless 
technique is that we cannot measure the radial velocities of weak and/or 
smoothly distributed [O~III] emission. Thus, long-slit spectra will often show 
a cleaner trend in the kinematics with position (e.g., compare the slitless 
measurements of Mrk 3 with the long-slit data in Ruiz et al. 2001). 
Nevertheless, we see clear trends in the radial velocities and FWHMs of 
emission-line knots as a function of projected radial distance from the optical 
continuum peaks in most of our objects. Trends as functions of distance or 
angle from the major axis of the NLR emission are weaker, if present, and 
likely require multiple long-slit observations to discern.

Most of the Seyfert 2 galaxies in our sample, along with NGC 1068 (Crenshaw \& 
Kraemer 2000) show significant blueshifts and redshifts on either side of their 
nuclei, which rules out simple rotation and is consistent with biconical 
outflow and the unified model, such that their bicone axes are relatively 
close to the plane of the sky. Our observations of the two Seyfert 1 galaxies 
NGC~3516 and NGC~4151 (Crenshaw et al. 2000) are also consistent with this 
picture; their elongated NLRs and the presence of blueshifts on one side of 
their nuclei and redshifts on the other indicate their bicone axes are at 
intermediate angles $(\sim$45\deg) with respect to our lines 
of sight. The Seyfert 2 galaxy NGC 5728 appears to be viewed at an 
intermediate angle in the biconical outflow picture, but not enough for us to
view the BLR directly.

Most of the Seyfert galaxies in our sample show a trend of increasing radial 
velocity to a projected angular distance of 0\arcsecpoint3 -- 2$''$, followed 
by a decrease to the systemic velocity further out. The radial acceleration 
could be due to radiation pressure or winds of more highly-ionized, diffuse 
gas. The source of the apparent deceleration is unknown, although one 
possibility is collision with an ambient medium (Crenshaw \& Kraemer 2000). The 
velocities in the ENLR are consistent with the finding that galactic rotation 
dominates the observed velocities (Unger et al. 1987), although the scatter is 
large, and we cannot rule out a secondary component of motion (such as radial 
outflow).

An interesting result from our slitless spectroscopy is that the emission-line 
knots, which have projected sizes of tens of pcs, are moving coherently, often 
with their own peculiar velocities. If they are indeed moving radially outward 
from the nucleus, their travel times out to several hundred pcs is on the order 
of 10$^{5}$ to 10$^{6}$ yr. It is interesting that the knots are able to 
maintain their integrity for this period of time, particularly since most have 
velocity dispersions on the order of several hundred km s$^{-1}$. The origin
of the high dispersions (e.g., expansion, turbulence) is not known, and the 
reason for the decrease in FWHM with position is also unknown.

Finally, it is interesting that 10 out of 12 Seyfert galaxies that we have 
studied so far have a very bright, resolved knot of [O III] emission in their 
inner NLR, approximately centered on the optical continuum peak. These nuclear 
knots of emission also have large FWHMs, on the order of the velocities found 
in the intrinsic UV absorbers in Seyfert 1 galaxies (Crenshaw et al. 1999). 
A large fraction, and possibly most, of the UV absorbers are thought to be 
located at distance of tens of pcs from their central AGN (Crenshaw, Kraemer, 
\& George 2003), and it is tempting to associate them with the nuclear knots of 
emission. Further work is needed to test this possible association.

\acknowledgments

Support for proposal 8340 was provided by NASA through a grant from the Space 
Telescope Science Institute, which is operated by the Association of 
Universities for Research in Astronomy, Inc., under NASA contract NAS 5-26555. 
Some of the data presented in this paper were obtained from the Multimission 
Archive at the Space Telescope Science Institute (MAST).

\clearpage

\figcaption[f1.eps]{WFPC2 emission-line and continuum images and STIS slitless 
spectrum of the [O~III] $\lambda$5007 emission line in Mrk 3. The WFPC2 images 
have the same spatial scale and orientation as the STIS image. The ``+'' sign 
gives the location of the opical continuum peak.}

\figcaption[f2.eps]{{\it HST} emission-line images and STIS slitless spectra of 
the [O~III] $\lambda$5007 emission line in nine Seyfert galaxies. The WFPC2 
images have the same spatial scale and orientation as the STIS image. The ``+'' 
signs give the locations of the opical continuum peaks. The diagonal bands in 
the NGC~5506 [O~III] image are artifacts of the image rotation.}

\figcaption[f3.eps]{Radial velocity and FWHM of the [O~III] emission as a 
function of angular distance from the optical continuum peak in Mrk 3.
Negative and positive positions are below and above the central continuum peak 
in the image, respectively.}

\figcaption[f4.eps]{Radial velocity and FWHM of the [O~III] emission as a 
function of angular distance from the optical continuum peak. Negative and 
positive positions are below and above the central continuum peaks in the 
images, respectively}

\figcaption[f5.eps]{Extracted spectrum from the nuclear [O~III] knot in NGC 
3081, plotted as a function of radial velocity relative to the systemic 
velocity of the host galaxy.}

\clearpage

\begin{deluxetable}{lcccccc}
\tablecolumns{7}
\footnotesize
\tablecaption{Sample \label{tbl-7}}
\tablewidth{0pt}
\tablehead{\colhead{Object} & \colhead{Seyfert} & 
\colhead{Morph.}
& \colhead{Redshift (cz)} &  \colhead{Redshift}  &  
\colhead{Redshift} & \colhead{ENLR Shift\tablenotemark{c}} \\  
\colhead{} & \colhead{Type\tablenotemark{a}}  & \colhead{Type}
& \colhead{(km s$^{-1}$)} &  \colhead{Source}  &  
\colhead{Reference\tablenotemark{b}} &  \colhead{(km s$^{-1}$)}
}
\startdata
Mrk 3    &  2   & S0     & 4050$\pm$6   &  21 cm H I  &  TC  &  $-$116 \\
Mrk 573  &  2   & SB0    & 5150$\pm$11  &  Stellar    &  NW  &  $+$195  \\
Mrk 620  &  2   & SB(r)a & 1840$\pm$4   &  21 cm H I  &  3RC     & 0 \\
NGC 1386 &  2   & SB0    & 918$\pm$34   &  21 cm H I  &  HR &  $-$88  \\
NGC 3081 &  2   & SB0    & 2367$\pm$9   &  21 cm H I  &  3RC & $+$92   \\
NGC 3516 &  1 & SB0    & 2503$\pm$75  &  21 cm H I  &  B   & $-$66    \\
NGC 5506 &  1.9 & Sa      & 1815$\pm$9   &  21 cm H I  &  3RC & 0 \\
NGC 5643 &  2   & SABc    & 1199$\pm$5   &  21 cm H I  &  3RC &  0 \\
NGC 5728 &  2   & SAB(r)a & 2788$\pm$8   &  21 cm H I  &  3RC &  0\\
NGC 7212 &  2   & Sa      & 8011$\pm$21  &  Stellar   &   NW  & $+$102 \\
\enddata
\tablenotetext{a}{Morphological type from de Vaucouleurs et al. 1991}
\tablenotetext{b}{TC- Tifft \& Cooke (1988), 
3RC- de Vaucouleurs et al. (1991), NW - Nelson \& Whittle (1995), 
HR - Huchtmeier \& Richter (1989), B - Balkowski et al. (1972)}
\tablenotetext{c}{Correction added to the radial velocities to force the average
ENLR velocity to be zero (see \S2.3.)}
\end{deluxetable}

\begin{deluxetable}{lccccc}
\tablecolumns{6}
\scriptsize
\tablecaption{STIS Observations of Seyfert Galaxies \label{tbl-2}}
\tablewidth{0pt}
\tablehead{\colhead{Galaxy} & \colhead{STScI}
& \colhead{Date} & \colhead{Central $\lambda$} 
& \colhead{Exposure} & \colhead{PA}\\
\colhead{} & \colhead{Dataset}
& \colhead{(UT)} & \colhead{(\AA)} 
& \colhead{(s)} & \colhead{(\deg)}}
\startdata
Mrk 3    & O5F403020  &  2000 Jan 16    & 5093 & 2154 &  112\\ 
Mrk 573  &  O5F401020  &  2000 Oct 24   & 5093 & 2096 &  330\\   
Mrk 620  &  O5F404020  &  2000 Mar 3 & 4961 & 2128 &  50\\   
NGC 1386 &  O5F402020  &  2000 Jun 23   & 4961 & 2106 &  175\\  
NGC 3081 & O5F405020  &  2000 Feb 15    & 5093 & 2108 &  335\\   
NGC 3516 & O5F406020  &  2000 Jan 18    & 5093 & 2154 &  205\\ 
NGC 5506 &  O5F407020  &  2000 Mar 18   & 4961 & 2096 &  206\\ 
NGC 5643 & O5F408020  &  2000 Feb 23    & 4961 & 2107 &  260\\ 
NGC 5728 &  O5F409020  &  2000 Apr 24   & 5093 & 2110 &  280\\  
NGC 7212 &  O5F410020 &  2000 Jun 4   & 5093 & 2112 &  185\\  
\enddata
\end{deluxetable}

\begin{deluxetable}{lcccccc}
\tablecolumns{7}
\footnotesize
\tablecaption{$HST$ Archival Images\label{tbl-3}}
\tablewidth{0pt}
\tablehead{\colhead{Galaxy} & \colhead{STScI} & \colhead{Date} 
& \colhead{Camera} & \colhead{Filter} & \colhead{Exposure} & 
\colhead{PA} \\
\colhead{} &  \colhead{Dataset} &  \colhead{(UT)} &  \colhead{} &
\colhead{} &  \colhead{(s)} &  \colhead{(\deg)}
}
\startdata
Mrk 3  &  W0MW0601T  &   1991 July 18  &   PC   &  F502N  &  1800 &  151\\
       &  X14W0301T  &   1992 Dec 11   &   FOC  &  F501N  &  1197 &  344 \\
       &  W0MW0602T  &   1991 July 18  &   PC   &  F547M  &  350  &  151 \\
       &  U2E62A01T  &  1994 Oct 20   &  WFPC2  &  F606W  &  500 &  298 \\
Mrk 573  &  U2XI0701T  &  1995 Nov 12  &  WFPC2 &  FR533N  &  300 & 214  \\
  &  X2580504T  &  1994 Feb 9  &   FOC   &  F550M   &   1300.9 & 39 \\
Mrk 620  &  U3A00301T  &  1997 Feb 5  &  WFPC2  &  FR533N   &  800  & 267 \\
  &  U3A00303T  &  1997 Feb 5  &  WFPC2  &  F547M    &  100  & 267 \\
NGC 1386 &  U3A00201M  &  1997 June 28 &  WFPC2 &  F502N  &  400  & 266\\ 
  &  U3A00203M  &  1997 June 28 &  WFPC2 &  F547M  &   40  & 266 \\ 
NGC 3081 &  U3A00501T  &  1997 Feb 5  &   WFPC2  &  FR533N  &  200 & 119 \\
  &  U3A00503T  &  1997 Feb 5  &   WFPC2  &  F547M   &  40  & 119 \\ 
NGC 3516 &  U3A00602T  &  1997 Feb 7  &   WFPC2  &  FR533N  &  800 & 343 \\
  &  U3A00603T  &  1997 Feb 7  &   WFPC2  &  F547M   &  70  & 343\\
NGC 5506 &  X2740302T  &  1995 Jan 29  &  FOC  &  F501N  &  1797   & 255\\
  &  U2E62001T  &  1994 July 21 &  WFPC2  &  F606W  &  500  & 155\\
  &  X2740301T  &  1995 Jan 29  &  FOC  &  F550M   &  1996  & 255\\
NGC 5643 &  U2NP0401T  &  1995 June 2  &  WFPC2  &  F502N   &  350 & 106 \\ 
  &  U2NP0403T  &  1995 June 2  &  WFPC2  &  F547M   &  50  & 106 \\ 
NGC 5728 &  W1150401T  &  1992 Sept 4  &  PC  &   F492M   &  600  & 288\\
  &  W1150403T  &  1992 Sept 4  &  PC  &   F547M   &  600  & 288\\
NGC 7212 &  U2XI0401T  &  1995 Sept 26 &  WFPC2 &  FR533N  &  300 & 248 \\
         &  U2XI0403T  &  1995 Sept 26 &  WFPC2 &  FR533N  &  140 & 248 \\
  &  U2E65301T  &  1994 Sept 27 &  WFPC2 &  F606W  &  500  & 160\\
\enddata
\end{deluxetable}
 

\clearpage
\begin{figure}
\epsscale{0.9}
\plotone{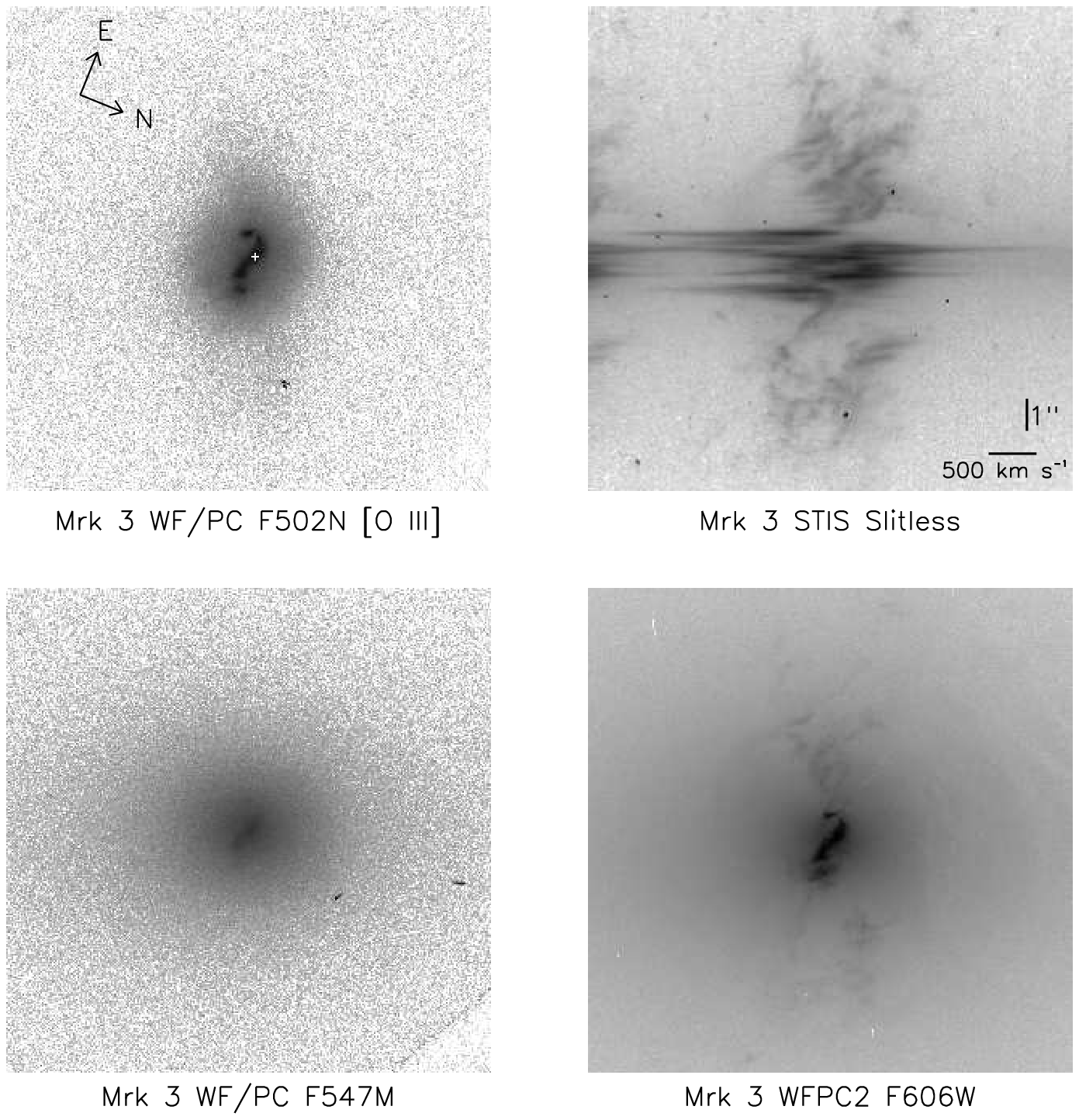}
\\Fig.~1.
\end{figure}

\clearpage
\begin{figure}
\plotone{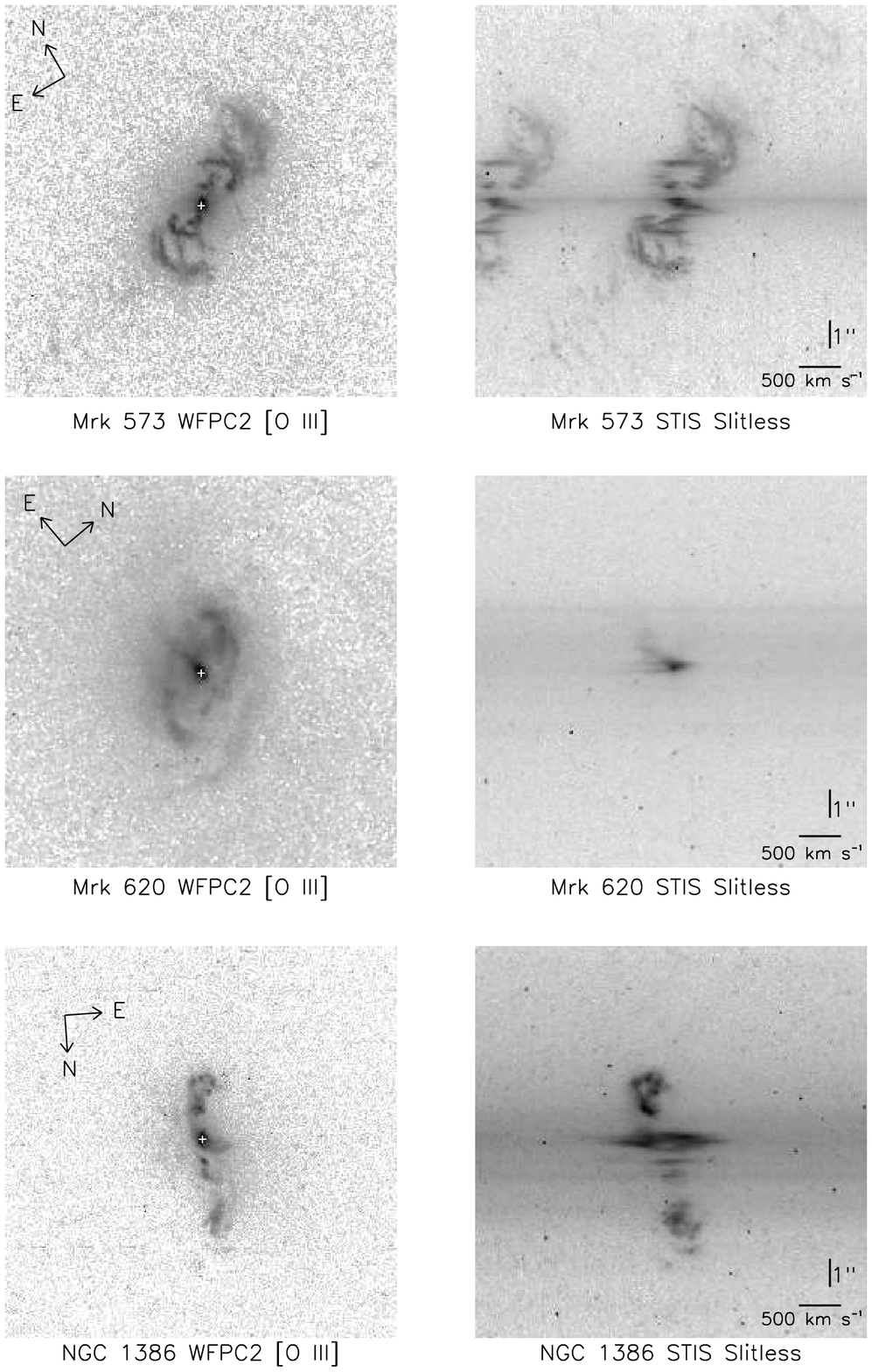}
\\Fig.~2a.
\end{figure}

\clearpage
\begin{figure}
\plotone{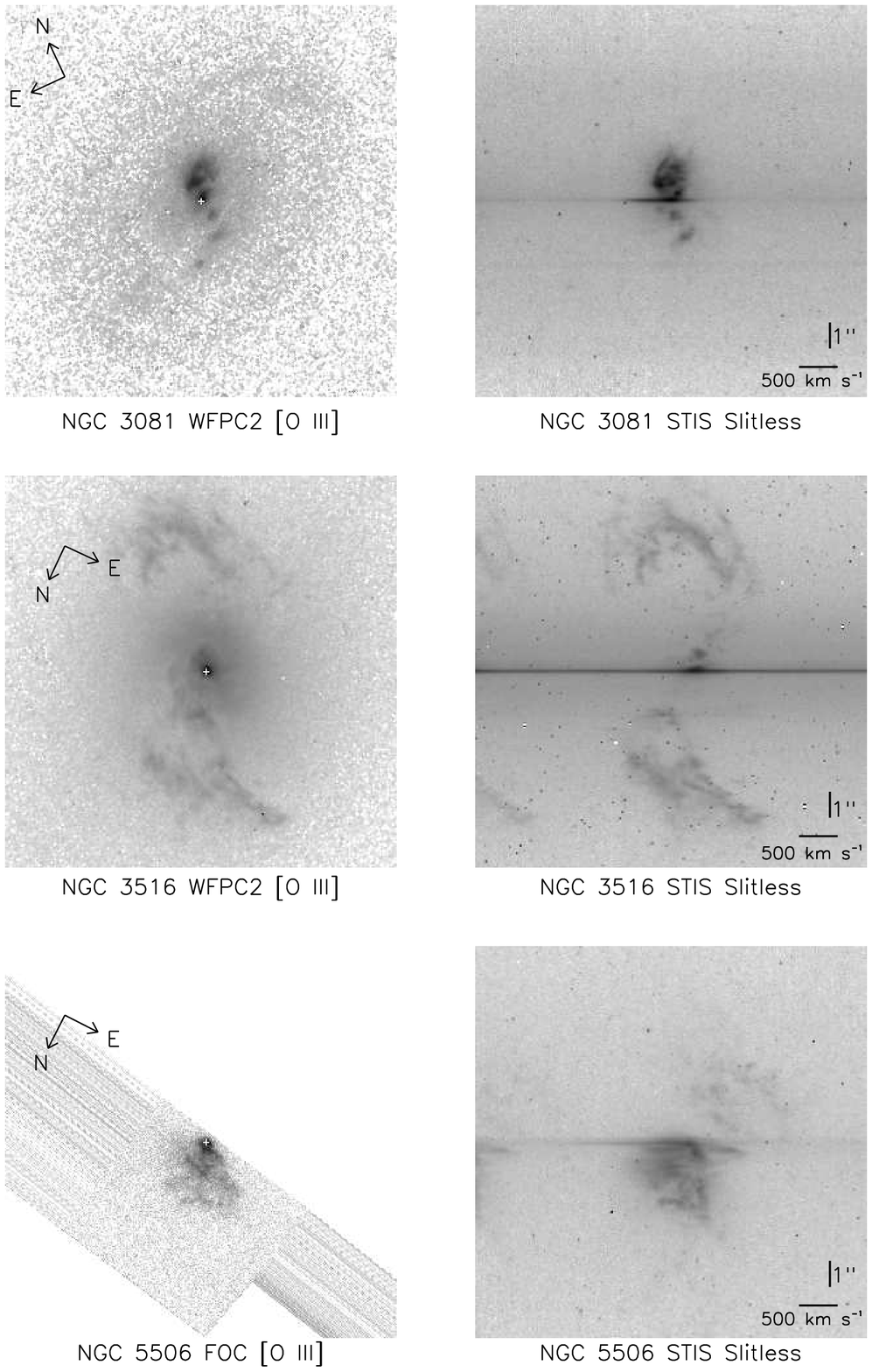}
\\Fig.~2b.
\end{figure}

\clearpage
\begin{figure}
\plotone{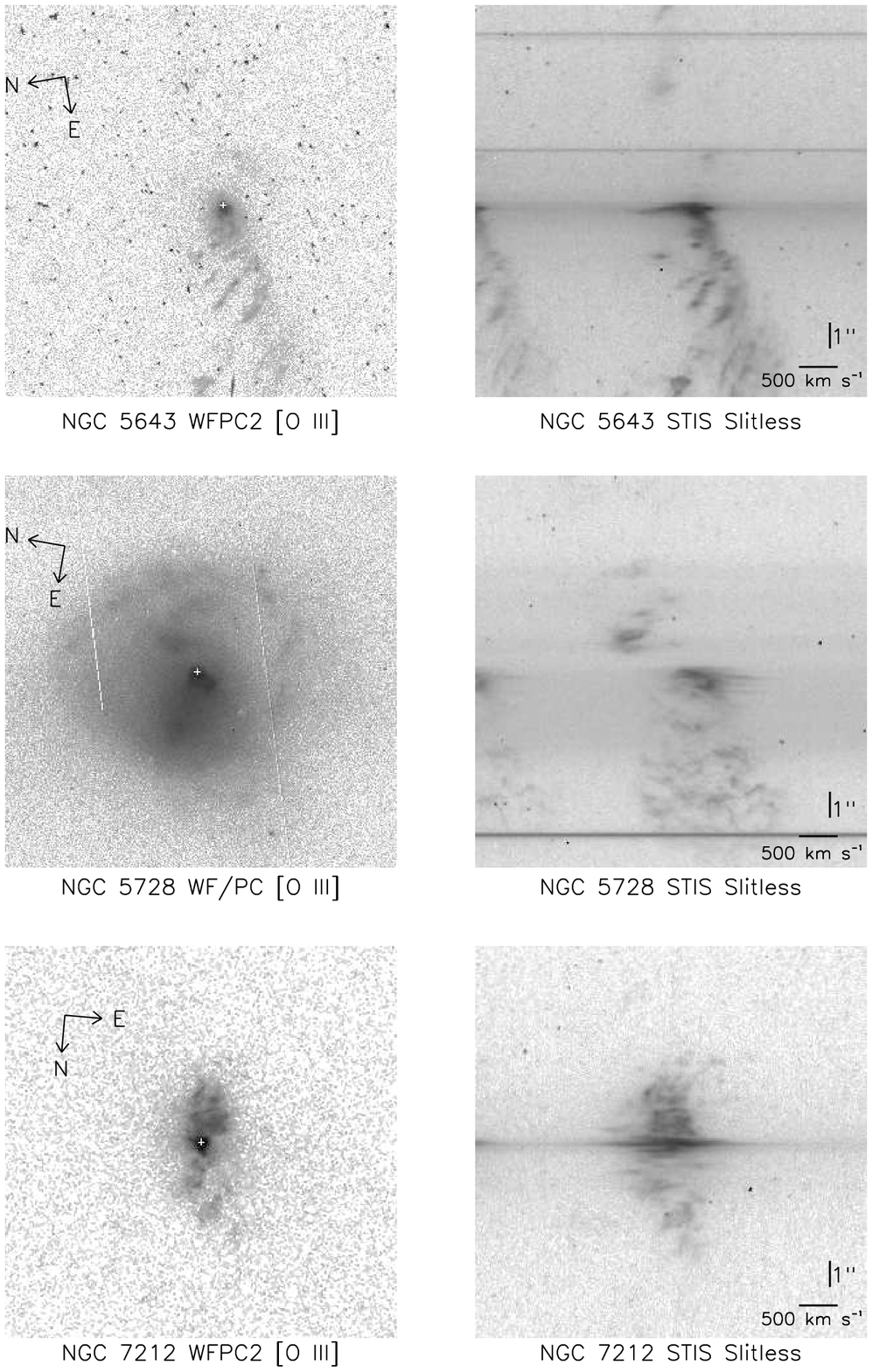}
\\Fig.~2c.
\end{figure}

\clearpage
\begin{figure}
\plotone{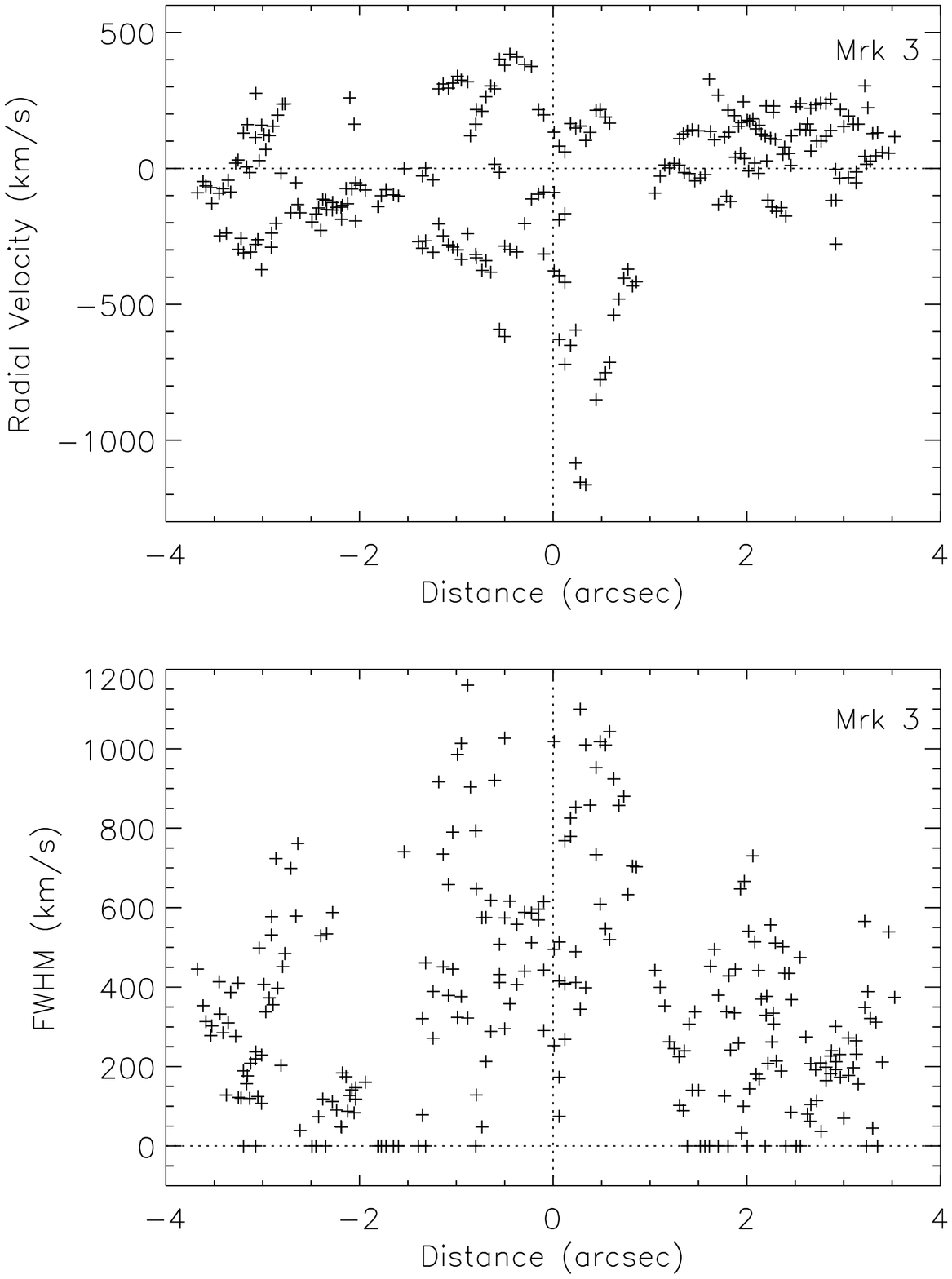}
\\Fig.~3.
\end{figure}

\clearpage
\begin{figure}
\plotone{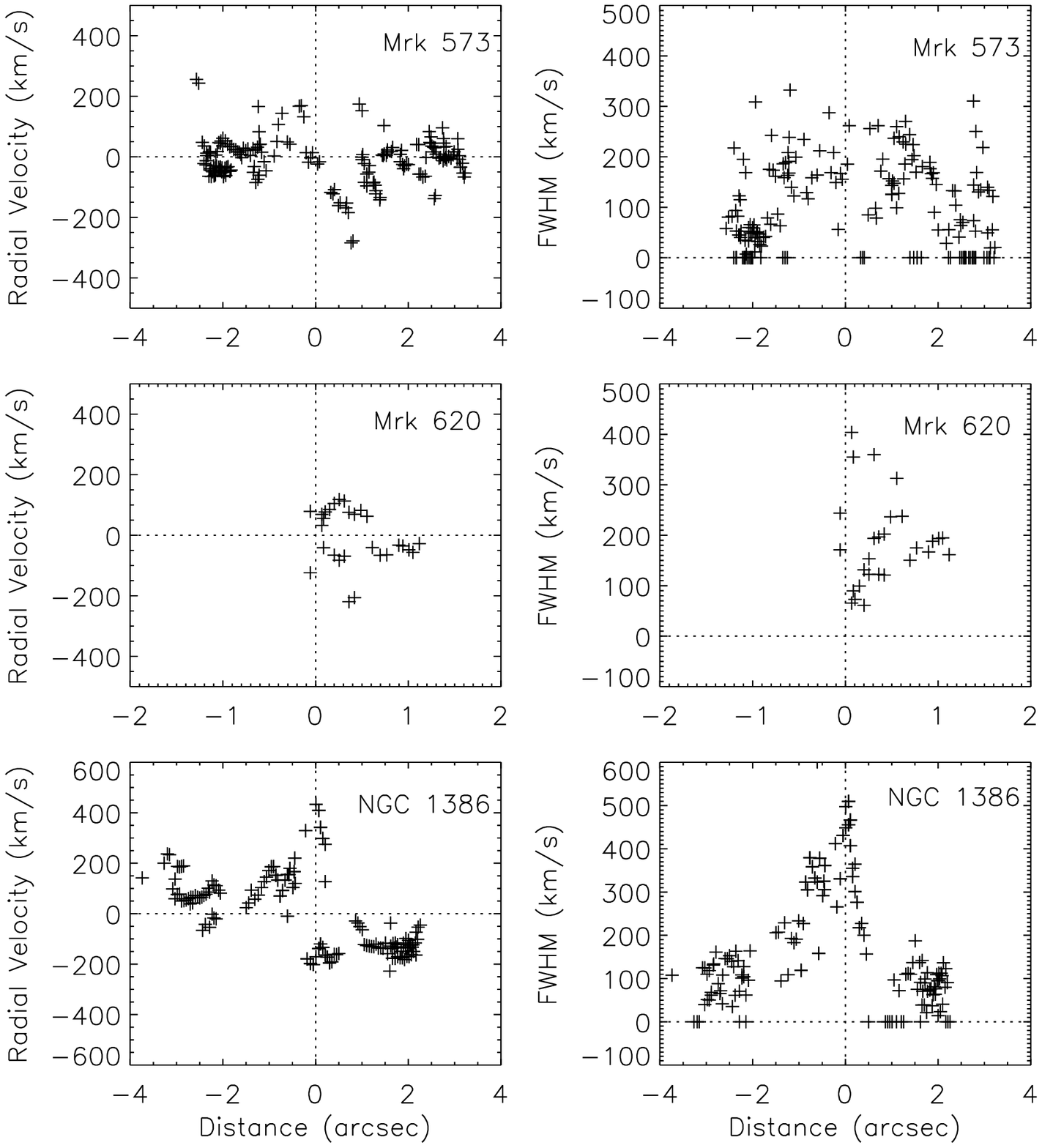}
\\Fig.~4a.
\end{figure}

\clearpage
\begin{figure}
\plotone{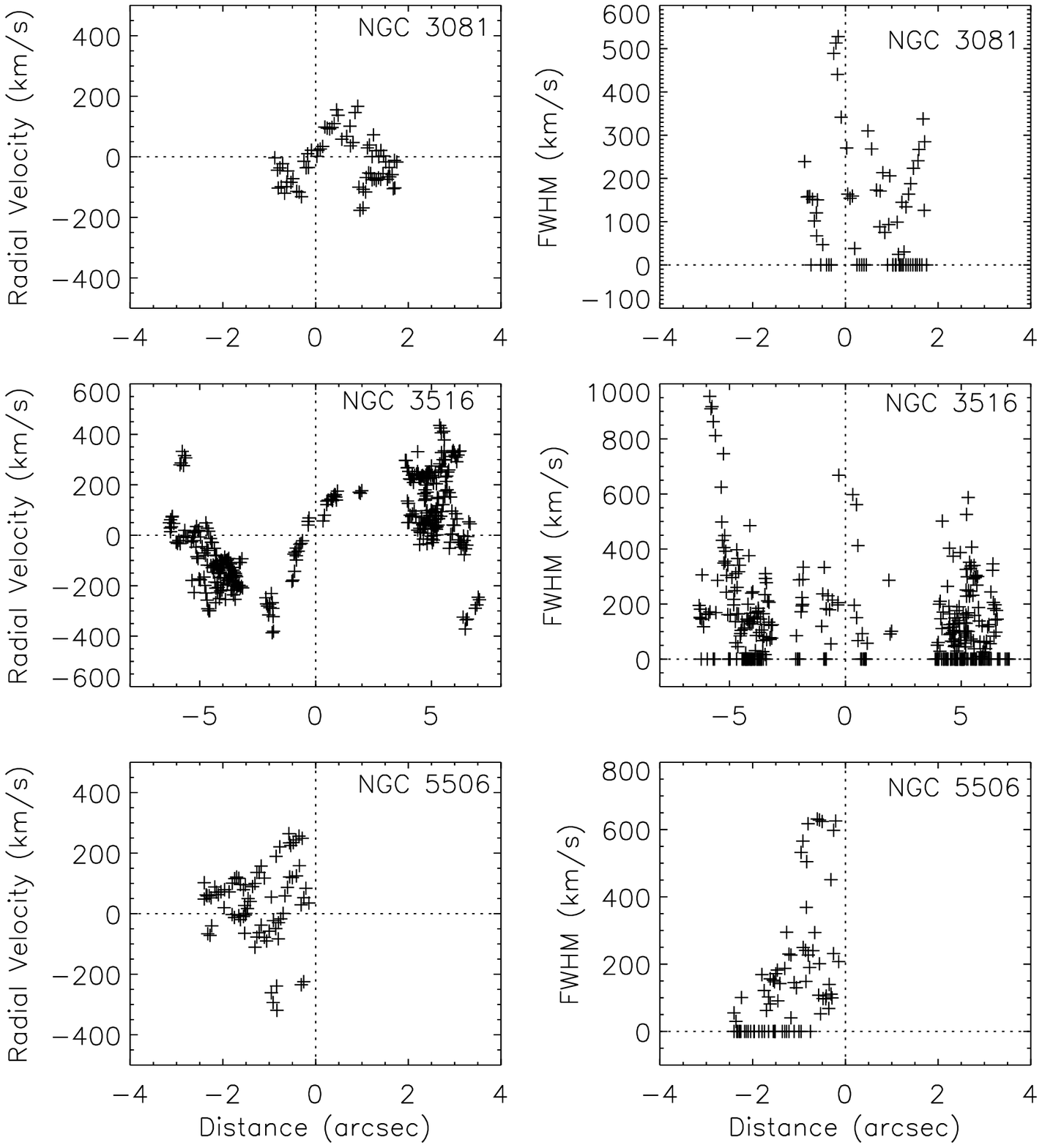}
\\Fig.~4b.
\end{figure}

\clearpage
\begin{figure}
\plotone{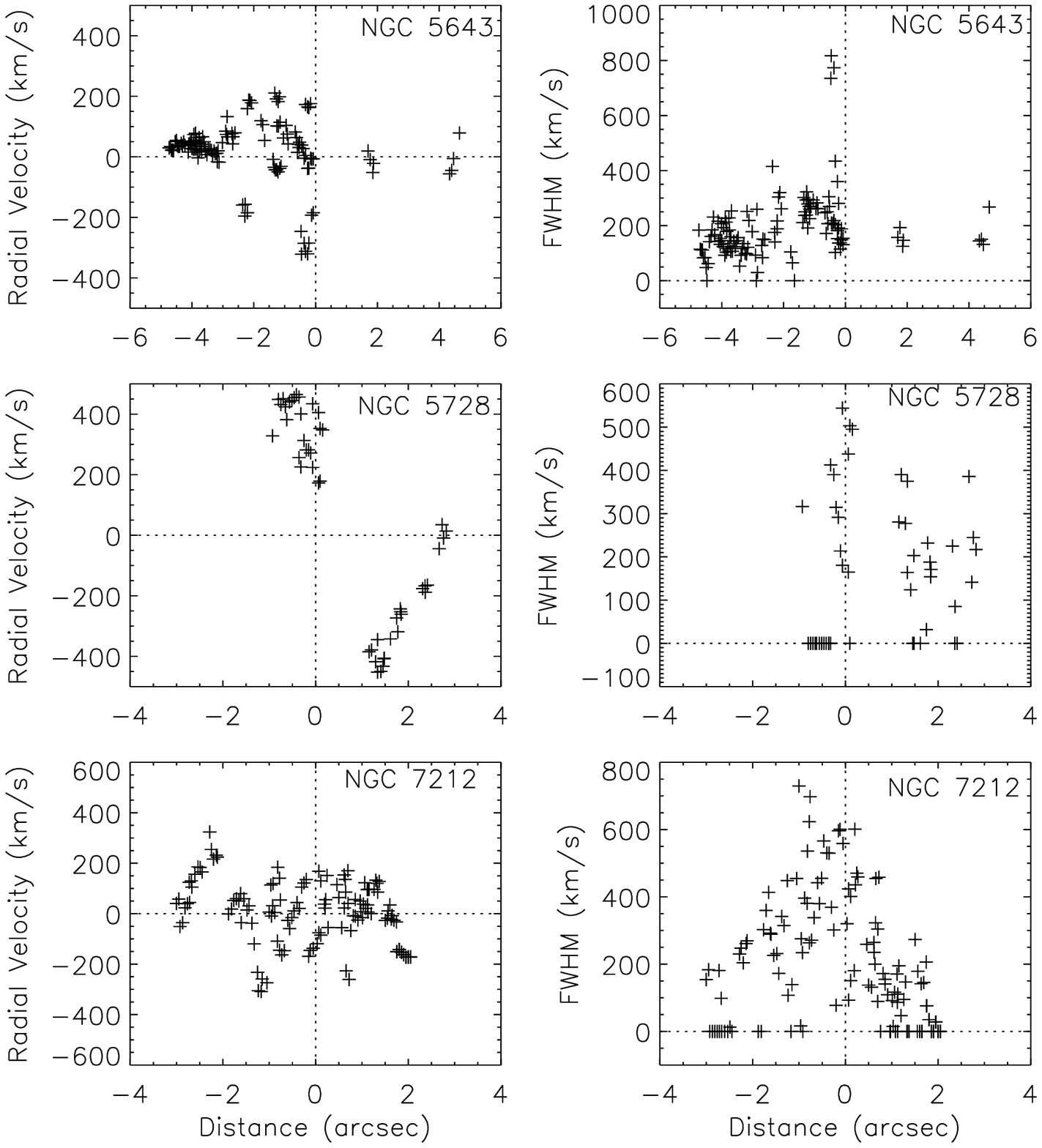}
\\Fig.~4c.
\end{figure}

\clearpage
\begin{figure}
\plotone{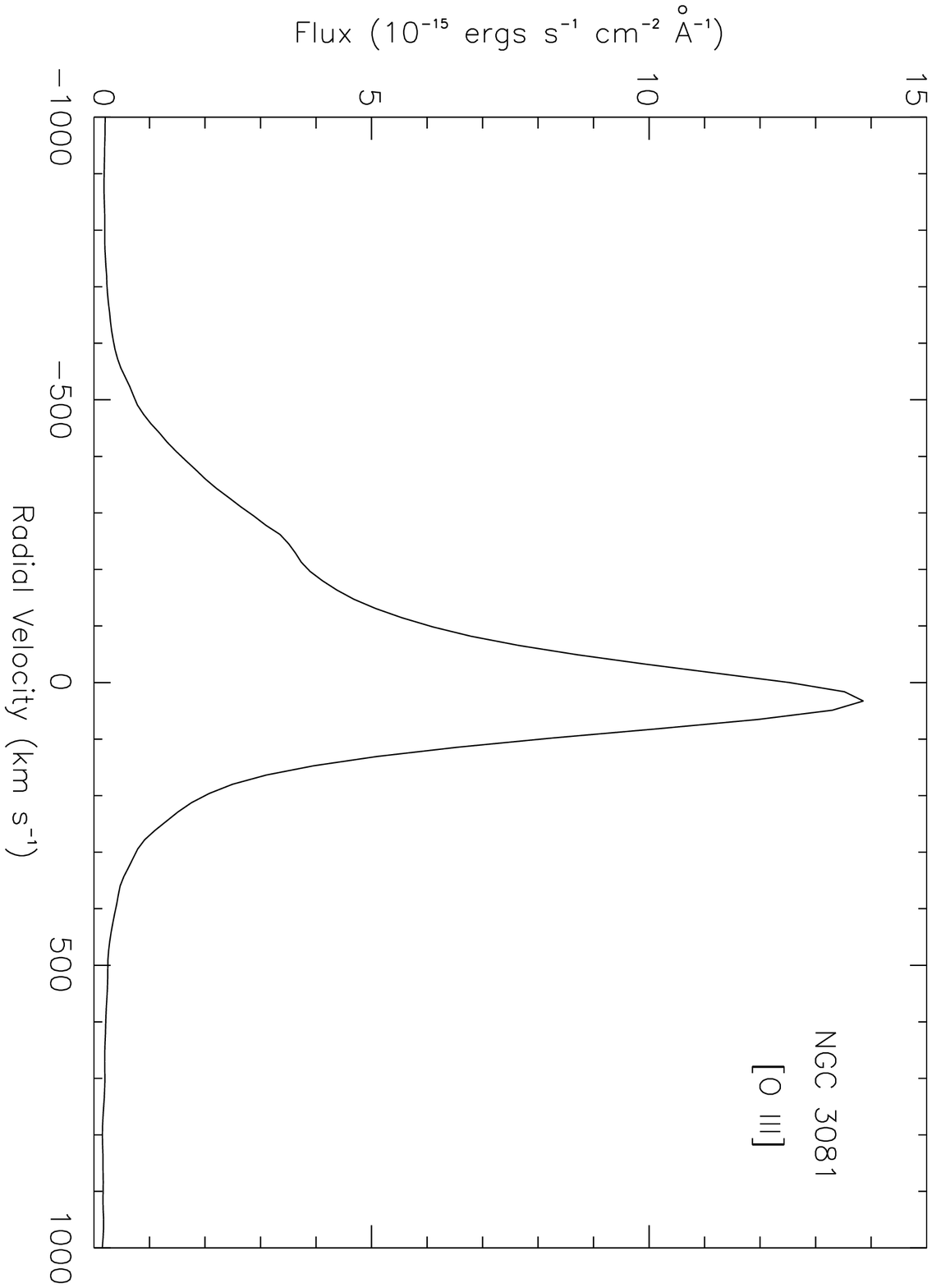}
\\Fig.~5.
\end{figure}

\end{document}